\documentclass[twoside]{article}
\usepackage{fleqn,espcrc1}

\usepackage{graphicx}

\newcommand{\nuc}[2]{\ensuremath{\mathrm {^{#2}#1}}}
\newcommand{\gk}{\ensuremath{\mathrm{\thinspace GK}}}

\title{Toward a Standard Model of Core Collapse Supernovae}

\author{A. Mezzacappa
\address{Physics Division\\
Oak Ridge National Laboratory\\
Bldg. 6010, MS 6354\\
P.O. Box 2008\\
Oak Ridge, TN 37831-6354}%
       }
       
\begin{document}

\maketitle

\begin{abstract}
In this paper, we discuss the current status of core collapse 
supernova models and the future developments needed to achieve  
significant advances in understanding the supernova mechanism 
and supernova phenomenology, i.e., in developing a supernova
standard model. 
\end{abstract}

\section{Introduction}
Beginning with the first numerical simulations conducted by Colgate and 
White\cite{cw66}, three decades of supernova modeling have established 
a basic supernova paradigm. The supernova shock wave---formed when the 
iron core of a massive star collapses gravitationally and rebounds as 
the core matter exceeds nuclear densities and becomes incompressible---stalls 
in the iron core as a result of enervating losses to nuclear dissociation 
and neutrinos. The failure of this ``prompt'' supernova mechanism sets the 
stage for a ``delayed'' mechanism, whereby the shock is reenergized by 
the intense neutrino flux emerging from the neutrinospheres carrying off 
the binding energy of the proto-neutron star\cite{w85,bw85}. This heating 
is mediated primarily by the absorption of electron neutrinos and antineutrinos 
on the dissociation-liberated nucleons behind the shock. This past decade 
has also seen the emergence of multidimensional supernova models, which 
have investigated the role convection, rotation, and magnetic fields 
may play in the explosion
\cite{hbhfc94,bhf95,jm96,mcbbgsu98a,mcbbgsu98b,fh99,khowc99}, in some 
cases invoking new explosion paradigms.

Although a plausible framework is now in place, fundamental questions 
about the explosion mechanism remain: Is there more than one mechanism? 
For example, are some core collapse supernovae neutrino driven while 
others driven by magnetohydrodynamic jets? Is there one mechanism for 
one class of massive stars---e.g., stars with progenitor masses between 
10 and 20 M$_{\odot}$---and another for another class of massive 
stars---e.g., more massive progenitors? Or 
is the mechanism tailored to each progenitor? For neutrino-driven 
supernovae, is the neutrino heating sufficient, or are multidimensional 
effects such as convection and rotation necessary? In addition, any
``standard model'' for core collapse supernovae must reproduce 
observed supernova phenomenology, such as neutron star kicks\cite{fbb98} 
and the polarization of supernova emitted light\cite{w99}.
To answer these questions and to develop a ``standard model,'' 
simulations in one, two, and three dimensions must be 
performed and coordinated.

\section{One-Dimensional Supernova Models}
The neutrino energy deposition behind the shock depends sensitively 
not only on the neutrino luminosities but also on the neutrino spectra 
and angular distributions in the postshock region, necessitating exact
multigroup Boltzmann neutrino transport near and above the 
neutrinospheres or a very good approximation of it. 
[Multigroup, i.e., multi-neutrino energy, transport is
necessary because the neutrino opacities are strongly energy dependent
and low- and high-energy neutrinos may be transported in very different
ways (e.g., diffusion versus free streaming) at any given spatial 
point in the core at any given time.]
Ten percent variations in any of these quantities can make the 
difference between explosion and failure in supernova models\cite{jm96,bg93}. 
Past simulations have implemented increasingly sophisticated approximations 
to multigroup Boltzmann transport, the most sophisticated of which is 
multigroup flux-limited diffusion\cite{br93,wm93}. 
A generic feature of this approximation is that it 
underestimates the isotropy of the neutrino angular distributions in the 
heating region and, thus, the heating rate\cite{ja92,mmbg98}. 
Therefore, the
question arises whether or not failures to produce explosions 
in past one-dimensional models were the result of 
the transport approximations employed. It is important 
to note that, without invoking proto-neutron star (e.g., neutron finger) 
convection, simulations that implement multigroup flux-limited diffusion 
do not produce explosions\cite{br93,wm93} (as we will 
discuss, the existence and vigor of 
proto-neutron star convection is a matter of 
debate\cite{mcbbgsu98a,bd96,kjm96}).

To begin to address the question posed above, we have been simulating the core 
collapse, bounce, and postbounce evolution of 13, 15, and 20 M$_{\odot}$ stars, 
beginning with the precollapse models of Nomoto and Hashimoto\cite{nh88}, with 
a new neutrino radiation hydrodynamics code for both Newtonian and general 
relativistic spherically symmetric flows: AGILE--BOLTZTRAN. BOLTZTRAN is a 
three-flavor Boltzmann neutrino transport solver\cite{mb93b,mm99}, now extended 
to fully general relativistic flows\cite{l00}. In the 13 M$_{\odot}$ Newtonian 
(gravity) simulation we present here, it is employed in the $O(v/c)$ limit. 
AGILE is a conservative, adaptive mesh, general relativistic hydrodynamics 
code\cite{l00,lt98}. 
Its adaptivity enables us to resolve and seamlessly follow the shock through the 
iron core into the outer stellar layers.

The equation of state of Lattimer and Swesty\cite{ls91} (LS
EOS) is employed to calculate the local thermodynamic state 
and nuclear composition of
the matter in nuclear statistical equilibrium (NSE).
For matter initially in the silicon layer, the
temperatures are insufficient to achieve NSE. In this
region, the radiation and electron components of the LS EOS are used, while an
ideal gas of \nuc{Si}{28} is assumed for the nuclear component. For typical 
hydrodynamic timesteps ($\sim .1$
millisecond), silicon burning occurs within a single timestep for T
$\sim 5 \gk$\cite{ht99}; therefore, when a fluid element exceeds a
temperature of 5 \gk\ in our simulation, the silicon is instantaneously burned, 
achieving NSE
and releasing thermal energy equal to the difference in nuclear binding
energy between \nuc{Si}{28} and the composition determined by the LS EOS.

Figure 1, taken from the simulation of Mezzacappa et 
al.\cite{mlmhtb00}, shows the radius-versus-time trajectories of equal mass 
(0.01M$_{\odot}$) shells in the stellar iron core and silicon layer
in a Newtonian simulation initiated from the 13 M$_{\odot}$ 
progenitor. 
Core bounce and the formation and propagation of the initial bounce 
shock are evident. This shock becomes an accretion shock, decelerating 
the core material passing through it. At $\approx$ 125 ms after bounce, 
the accretion shock stalls at a radius $\approx$ 250 km and 
begins to recede, continuing to do so during the first 500 ms
of postbounce evolution. No explosion has developed in this model during 
this time. Similar behavior is exhibited in our 15 and 20 M$_{\odot}$ 
Newtonian models and in our 13 and 20 M$_{\odot}$ general relativistic
models, although these have not yet reached 500 ms after bounce. 
We will continue these
simulations and report on the final outcomes in subsequent papers.
If explosions are consistently obtained in the more realistic general 
relativistic cases, this would imply that core collapse supernovae
driven purely by neutrino heating are possible, although these same
models would have to be considered in three dimensions to assess 
whether or not multidimensional effects alter this conclusion and,
of course, to model any associated phenomenology, such as neutron 
star kicks.

\begin{figure}[h]
\begin{center}
\includegraphics[width=4.75in]{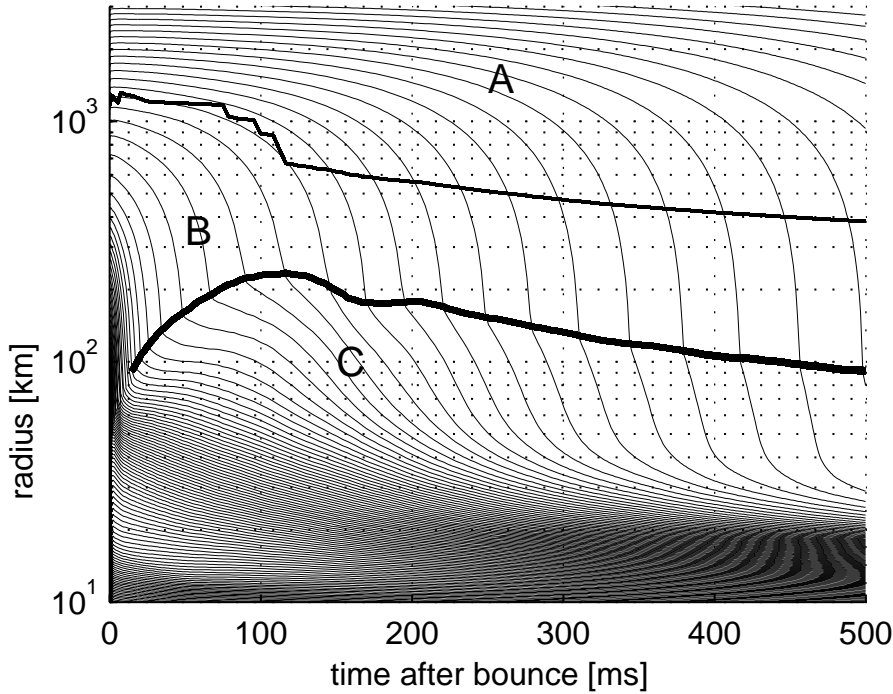}
\caption{We trace the shock, nuclear burning, and dissociation 
fronts (the shock and dissociation fronts are coincident), which carve out 
three regions in the $(r,t)$ plane. 
A: Silicon. 
B: Iron produced by infall compression and heating. 
C: Free nucleons and alpha particles.}
\end{center}
\end{figure}

\section{Convection}
Supernova convection falls into two categories: (1) convection 
near or below the neutrinospheres, which we refer to as proto-neutron
star convection, and (2) convection between the gain radius and the
shock, which we refer to as neutrino-driven convection. The gain
radius is the radius at which neutrino heating and cooling via
electron neutrino and antineutrino absorption and emission 
between the neutrinospheres and the shock balance. There is
net neutrino heating above this radius and net neutrino cooling
below it.

Proto-neutron star convection may aid the explosion mechanism by boosting 
the neutrinosphere luminosities.  Hot, lepton-rich 
rich matter is convectively transported 
to the neutrinospheres. This mode of convection may develop owing 
to instabilities caused by lepton and entropy gradients established by the 
deleptonization of the proto-neutron star via electron neutrino escape near 
the electron neutrinosphere and by the weakening supernova shock (as the 
shock weakens, it causes a smaller entropy jump in the material flowing 
through it). Proto-neutron star convection is arguably the most difficult 
to investigate numerically because the neutrinos and the matter are coupled 
and, consequently, multidimensional simulations must include both 
multidimensional hydrodynamics and multidimensional, multigroup neutrino 
transport.

Neutrino-driven convection may aid the explosion mechanism by boosting the 
shock radius and the neutrino heating efficiency, thereby facilitating shock 
revival. It develops as the result of the entropy gradient established 
as the shocked stellar core material infalls between the shock and the 
gain radius, being continually heated in the process.

\subsection{Proto-Neutron Star Convection (1D): Neutron Fingers}
The fundamental difficulty in modeling convection in spherically
symmetric models is apparent: convection is a three-dimensional
phenomenon, and spherically symmetric models can incorporate
convection only in a phenomenological way (e.g., via a mixing-length 
prescription). Moreover, because convection is not admitted by the 
one-dimensional hydrodynamics equations, some imposed criterion for the 
existence of convection must be used.

Neutron-finger convection has been invoked by Wilson et al.\cite{wm93}
in their one-dimensional models and has been deemed necessary by them to 
obtain supernova explosions. This mode of proto-neutron star convection
arises in the presence of a negative electron fraction gradient and a 
positive entropy gradient in the postshock stellar core, resulting in 
higher-entropy, neutron-richer matter above lower-entropy, neutron-poorer 
matter in the core. With the assumption that energy transport by neutrinos 
is more efficient than lepton transport, neutron fingers develop under 
these conditions, resulting (like salt fingers in the ocean) in
finger-like downflows of neutron-rich matter that penetrate deep into 
the stellar core. The assumption that energy transport is more efficient 
than lepton transport is justified in the following way: Three flavors 
of neutrinos (electron, muon, and tau) can transport energy, whereas 
only one (electron) can transport lepton number. However, detailed 
neutrino equilibration experiments carried out by Bruenn and Dineva\cite{bd96} 
demonstrate that the muon and tau neutrinos do not couple strongly with 
the stellar core matter in energy, and therefore, there is only one 
flavor (electron) that transports both energy and lepton number 
efficiently. Given the outcome of these numerical experiments, 
the fundamental assumption made by Wilson et al. should be reexamined. 
However, it is also important to note that the equilibration 
experiments carried out by Bruenn and Dineva have to be 
repeated in light of energy exchange channels between the
muon and tau neutrinos and the stellar core matter that have 
recently been identified (e.g., see the contribution by Thompson
and Burrows in this volume\cite{tb00}).

\subsection{Proto-Neutron Star Convection (2D): Ledoux Convection}
In certain regions of the stellar core, neutrino transport can equilibrate 
a convecting fluid element with its surroundings in both entropy and lepton 
number on time scales shorter than convection time scales, rendering 
the fluid element nonbouyant. This will occur in intermediate regimes 
in which neutrino transport is efficient but in which the neutrinos are still 
strongly enough coupled to the matter. Figures 2 and 3 from Mezzacappa 
et al.\cite{mcbbgsu98a} demonstrate that this equilibration can in fact 
occur. Figure 2 shows the onset and development of proto-neutron star 
convection in a 25 M$_{\odot}$ model shortly after bounce in a simulation 
that did not include neutrino transport, i.e., that was a hydrodynamics-only 
run. Figure 3 on the other hand shows the lack of any significant onset and 
development of convection when neutrino transport was included in what was 
otherwise an identical model. Transport's damping effects are obvious. (The 
same result occurred in our 15 M$_{\odot}$ model\cite{mcbbgsu98a}.) 

\begin{figure}[h]
\begin{center}
\includegraphics[width=4.75in]{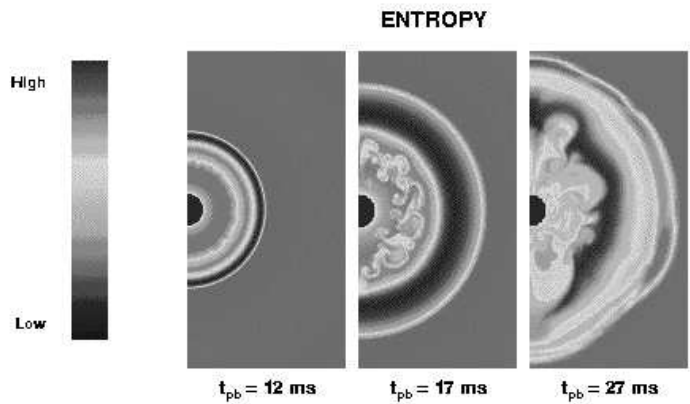}
\caption{Two-dimensional entropy plots showing the evolution of
proto-neutron star convection in our hydrodynamics-only 25 ${\rm M}_{\odot}$ 
model at 12, 17, and 27 ms after bounce.}
\end{center}
\end{figure}

\begin{figure}[h]
\begin{center}
\includegraphics[width=4.75in]{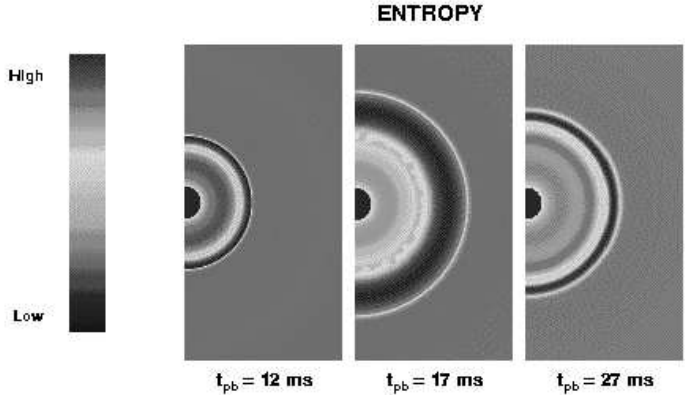}
\caption{ Two-dimensional entropy plots showing the evolution of
proto-neutron star convection in our hydrodynamics-plus-neutrino-transport 
25 ${\rm M}_{\odot}$ model at 12, 17, and 27 ms after bounce.}
\end{center}
\end{figure}

On the other hand, in the model of Keil et al.\cite{kjm96}, vigorous 
proto-neutron star convection developed, which then extended deep into 
the core as a deleptonization wave moved inward owing to neutrinos 
diffusing outward. In this model, convection occurs very deep in the 
core where neutrino opacities are high and transport becomes inefficient 
in equilibrating a fluid element with its surroundings. 

It is also important to note in this context that Mezzacappa et al.
and Keil et al. used complementary transport approximations. In
the former case, spherically symmetric transport was used, which
maximizes lateral neutrino transport and overestimates the 
neutrino--matter equilibration rate; in the latter case, ray-by-ray
transport was used, which minimizes (zeroes) lateral transport and
underestimates the neutrino--matter equilibration rate.

These outcomes clearly demonstrate that to determine whether or not 
proto-neutron star convection (including neutron fingers) exists and, 
if it exists, is vigorous, 
will require simulations coupling three-dimensional, multigroup neutrino 
transport and three-dimensional hydrodynamics. Moreover, realistic 
high-density neutrino opacities will also be needed. 

\subsection{Neutrino-Driven Convection (2D)}
This mode of convection occurs directly between the gain
radius and the stalled shock as a result of the entropy 
gradient that forms as material infalls between the 
two while being continually heated from below. In Figure 4, a sequence of
two-dimensional plots of entropy are shown, illustrating
the development and evolution of neutrino-driven convection
in our 15 M$_{\odot}$ model\cite{mcbbgsu98b}. High-entropy, rising
plumes and lower-entropy, denser, finger-like downflows are seen.
The shock is distorted by this convective activity.

\begin{figure}[h]
\begin{center}
\includegraphics[width=4.75in]{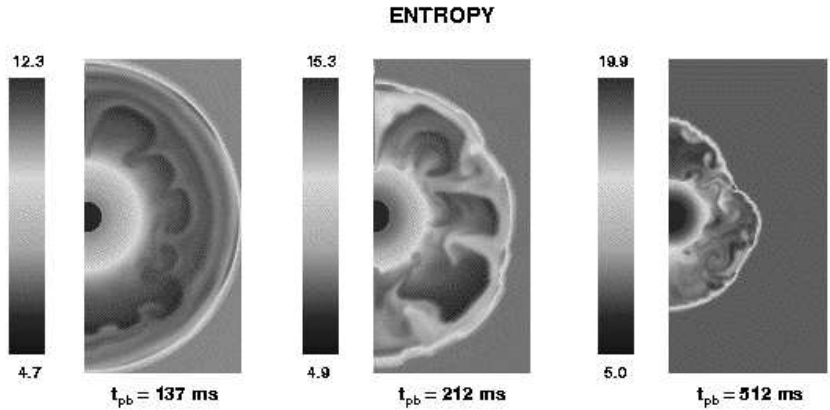}
\caption{Two-dimensional entropy plots showing the evolution of
neutrino-driven convection in our 15 ${\rm M}_{\odot}$ model
at 137, 212, and 512 ms after bounce.}
\end{center}
\end{figure}

In the Herant et al.\cite{hbhfc94} simulations, large-scale 
convection developed beneath the shock, leading to increased neutrino 
energy deposition, the accumulation of mass and energy in the gain 
region, and a thermodynamic engine they claimed ensured explosion, although
Herant et al. stressed the need for more sophisticated multidimensional, 
multigroup transport in future models. [They used two-dimensional ``gray''
(neutrino-energy--integrated, as opposed to multigroup) flux-limited
diffusion in neutrino-thick regions and a neutrino lightbulb approximation 
in neutrino-thin regions. In a lightbulb approximation, the neutrino 
luminosities and rms energies are assumed constant with radius.]
In the Burrows et al. simulations\cite{bhf95}, neutrino-driven convection 
in some models significantly boosted the shock radius and led to explosions. 
However, they stressed that success or failure in producing explosions was 
ultimately determined by the values chosen for the neutrino spectral parameters 
in their gray ray-by-ray (one-dimensional) neutrino diffusion scheme. [In 
spherical symmetry (1D), all rays are the same. In a ray-by-ray 
scheme in axisymmetry (2D), not 
all rays are the same, although the transport along each ray is a 1D problem. 
In this latter case, lateral transport between rays is ignored.] 
Focusing on the neutrino luminosities, Janka and M\"{u}ller\cite{jm96}, using 
an adjustable central neutrino lightbulb, conducted a parameter survey and 
concluded that neutrino-driven convection aids explosion only in a narrow 
luminosity window ($\pm 10\%$), below which the luminosities are too low 
to power explosions and above which neutrino-driven convection is not necessary
to power explosions. 
In more recent simulations carried out by Swesty\cite{s98} using two-dimensional 
gray flux-limited diffusion in both neutrino-thick and neutrino-thin regions, it 
was demonstrated that the simulation outcome varied dramatically as the 
matter--neutrino ``decoupling point,'' which in turn sets the neutrino spectra 
in the heating region, was varied within reasonable limits. (The 
fundamental problem in gray transport schemes is that the neutrino 
spectra, which are needed for the heating rate, are not computed. 
The spectra are specified by choosing a neutrino ``temperature,''
normally chosen to be the matter temperature at decoupling. In a
multigroup scheme, the spectra are computed dynamically.)  
In our two-dimensional models\cite{mcbbgsu98b}, the angle-averaged shock radii 
do not differ significantly from the shock trajectories in 
their one-dimensional counterparts, and no explosions are 
obtained, as seen in Figure 4. Neither the luminosities nor 
the neutrino spectra are free parameters. Our two-dimensional 
simulations implemented precomputed spherically symmetric (1D) multigroup 
flux-limited diffusion neutrino transport, compromising transport 
dimensionality to implement multigroup transport and a seamless 
transition between neutrino-thick and neutrino-thin regions, although 
without feedback between the hydrodynamics and the transport.

In light of the neutrino transport approximations made, the fact that 
none of the simulations have been three dimensional, and the mixed 
outcomes, next-generation simulations will have to reexplore neutrino-driven 
convection in the context of three-dimensional simulations that implement 
more realistic three-dimensional multigroup neutrino transport.

\section{Rotation and Magnetic Fields}
The observed polarization of spectra in a number of core collapse supernovae
is a fingerprint of gross asymmetries in the explosions, with axis ratios 
as large as 2:1, which has lead to the suggestion that these explosions must 
have been jet-like\cite{w99}. If so, what is the origin of these jet-like 
explosions? Are the jets formed early on as part of the explosion 
mechanism, or do they form as the supernova shock wave propagates 
through an aspherical medium? Three classes of models that exhibit 
jet-like behavior have been investigated: neutrino-driven jet models 
after core collapse and bounce in a rotating stellar core\cite{fh99}, 
MHD-driven jet models from the core collapse and bounce of rotating, 
magnetized cores\cite{lw70,sy84,khowc99}, and neutrino- or MHD-driven 
jet models around a newly formed black hole in a failed or successful 
supernova, respectively, in a rotating stellar core\cite{mw99}. Two 
things of note: (1) Having jet-like explosions does not imply the 
explosions are not neutrino-driven. (2) There may be several explosion 
paradigms at work in the full range of observed supernovae of all 
progenitor stars. 

In the two-dimensional axisymmetric model of Fryer and Heger\cite{fh99}, 
the first to include the effects of both convection and rotation, a weak, 
jet-like explosion along the rotation axis was obtained. Unfortunately, all 
axisymmetric simulations must contend with reflecting boundary conditions on 
the rotation axis, which numerically bias on-axis outflow, not allowing the 
flow to turnover and thereby break axisymmetry. Moreover, in three dimensions, 
the interaction of convection and rotation would result in ``tornadic'' flows 
not admitted by the imposed axisymmetry, which implies that the flow should be 
qualitatively
different in three dimensions. In the two-dimensional models of Leblanc and 
Wilson\cite{lw70} and Symbalisty\cite{sy84}, it was concluded that unusually 
high rotation rates and magnetic field strengths were required to
generate explosions, requiring, for example, magnetic field strengths 
as large as $10^{17}$ Gauss. In the model of MacFadyen and Woosley\cite{mw99},
although a plausible neutrino energy deposition mechanism was invoked,
the energy deposition was parameterized and not modeled in detail. 
Thus, a full exploration of these possibilities will require 
three-dimensional (neutrino) radiation hydrodynamics, magnetohydrodynamics, 
and magneto-radiation hydrodynamics simulations.

\section{Outlook}
Fundamental questions in supernova theory remain, which can only be 
answered via systematic and coordinated simulations in one, two, and 
three dimensions.

We have shown results from the first 500 ms of a 
one-dimensional (spherically symmetric) Newtonian 
simulation with Boltzmann neutrino transport 
initiated from a 13 M$_{\odot}$ progenitor. In light of our 
implementation of Boltzmann transport, if we do not obtain 
explosions in this model, or other models we have initiated from 
different progenitors (see also Rampp and Janka\cite{rj00}), it 
would suggest that improvements in our initial conditions 
(precollapse models) and/or input physics are needed, and/or that 
the inclusion of multidimensional effects such as convection, rotation, 
and magnetic fields are required ingredients in the recipe for explosion. 
In the past, it was not clear whether failure 
in spherically symmetric models was the result of transport approximations 
or the lack of inclusion of important physics. With the implementation of 
Boltzmann transport, this conclusion can be 
made unambiguously. We will report on the continued 
evolution of our 13 M$_{\odot}$ model and on our other 
models in subsequent papers.

Potential improvements in our initial conditions and input physics 
include: improvements 
in precollapse models\cite{ba98,unn99,hlw00,hlmw00}; the use of ensembles of 
nuclei in the stellar core rather than a single representative nucleus; 
computing the neutrino--nucleus cross sections with detailed shell model 
computations\cite{lmp00}; and the inclusion of nucleon correlations in the 
high-density neutrino opacities\cite{bs98,rplp99}. These improvements 
all have the potential 
to quantitatively, if not qualitatively, change the details of our simulations. 
Thus, it is important to note that the conclusions drawn here are drawn 
considering the initial conditions and input physics used.

To accurately investigate multdimensional effects such as convection, rotation,
and magnetic fields, future simulations must be carried out in three
dimensions and must implement realistic, three-dimensional, multigroup
neutrino transport. Three-dimensional simulations will be necessary to 
assess, for example, the vigor of convection in the proto-neutron star, 
where the neutrinos and the matter are strongly coupled and the flow is 
three-dimensional, and to assess the character of neutrino-driven convection 
behind the shock in a stellar core that is both rotating and convecting.
Certainly, three-dimensional simulations are required to study the 
development of MHD jets in stellar cores, and given that the development 
of such jets depends in some scenarios on the convection in the 
proto-neutron star, which in turn will depend on the neutrino transport, 
all of these studies must be strongly coupled.
 
\section*{Acknowledgments}
A.M. is supported at the Oak Ridge National Laboratory, managed by
UT-Battelle, LLC, for the U.S. Department of Energy under contract
DE-AC05-00OR22725. He would like to thank the organizers, Joergen 
Christensen-Dalsgaard and Karlheinz Langanke, for a delightful 
conference, and Raph Hix for valuable comments on the manuscript.


\begin{thebibliography}{10}

\bibitem{cw66}
S.~A. Colgate and R.~H. White, Astrophysical Journal {\bf 143},  626  (1966).

\bibitem{w85}
J.~R. Wilson,  in {\em Numerical Astrophysics}, edited by J.~M. Centrella,
  J.~M. LeBlanc, and R.~L. Bowers (Jones and Bartlett, Boston, 1985), pp.\
  422--434.

\bibitem{bw85}
H.~H. Bethe and J.~R. Wilson, Astrophysical Journal {\bf 295},  14  (1985).

\bibitem{hbhfc94}
M. Herant {\it et~al.}, Astrophysical Journal {\bf 435},  339  (1994).

\bibitem{bhf95}
A. Burrows, J. Hayes, and B.~A. Fryxell, Astrophysical Journal {\bf 450},  830
  (1995).

\bibitem{jm96}
H.-T. Janka and E. M\"{u}ller, Astronomy and Astrophysics {\bf 306},  167
  (1996).

\bibitem{mcbbgsu98a}
A. Mezzacappa {\it et~al.}, Astrophysical Journal {\bf 493},  848  (1998).

\bibitem{mcbbgsu98b}
A. Mezzacappa {\it et~al.}, Astrophysical Journal {\bf 495},  911  (1998).

\bibitem{fh99}
C.~L. Fryer and A. Heger, astro-ph/9907433  .

\bibitem{khowc99}
A.~M. Khokhlov {\it et~al.}, Astrophysical Journal {\bf 524},  L107  (1999).

\bibitem{fbb98}
C.~L. Fryer, A. Burrows, and W. Benz, Astrophysical Journal {\bf 496},  333
  (1998).

\bibitem{w99}
J.~C. Wheeler,  in {\em Cosmic Explosions, A Conference Proceedings}, edited by
  S.~S. Holt (AIP, Melville, 2000).

\bibitem{bg93}
A. Burrows and J. Goshy, Astrophysical Journal Letters {\bf 416},  L75  (1993).

\bibitem{br93}
S.~W. Bruenn,  in {\em First Symposium on Nuclear Physics in the Universe},
  edited by M.~W. Guidry and M.~R. Strayer (IOP Publishing, Bristol, 1993),
  pp.\ 31--50.

\bibitem{wm93}
J.~R. Wilson and R.~W. Mayle, Physics Reports {\bf 227},  97  (1993).

\bibitem{ja92}
H.-T. Janka, Astronomy and Astrophysics {\bf 256},  452  (1992).

\bibitem{mmbg98}
O.~E.~B. Messer, A. Mezzacappa, S.~W. Bruenn, and M.~W. Guidry, Astrophysical
  Journal {\bf 507},  353  (1998).

\bibitem{bd96}
S.~W. Bruenn and T. Dineva, Astrophysical Journal Letters {\bf 458},  L71
  (1996).

\bibitem{kjm96}
W. Keil, H.-T. Janka, and E. M\"{u}ller, Astrophysical Journal Letters {\bf
  473},  L111  (1996).

\bibitem{nh88}
K. Nomoto and M. Hashimoto, Physics Reports {\bf 163},  13  (1988).

\bibitem{mb93b}
A. Mezzacappa and S.~W. Bruenn, Astrophysical Journal {\bf 405},  669  (1993).

\bibitem{mm99}
A. Mezzacappa and O.~E.~B. Messer, Journal of Computational and Applied
  Mathematics {\bf 109},  281  (1998).

\bibitem{l00}
M. Liebend\"{o}rfer, Ph.D. thesis, University of Basel, Basel, Switzerland,
  2000.

\bibitem{lt98}
M. Liebend\"{o}rfer and F.-K. Thielemann,  in {\em Nineteenth Texas Symposium
  on Relativistic Astrophysics}, edited by E. Aubourg, T. Montmerle, J. Paul,
  and P. Peter (Elsevier Science B.~V., Amsterdam, 2000).

\bibitem{ls91}
J. Lattimer and F.~D. Swesty, Nuclear Physics {\bf A535},  331  (1991).

\bibitem{ht99}
W.~R. Hix and F.-K. Thielemann, Astrophysical Journal {\bf 511},  862  (1999).

\bibitem{mlmhtb00}
A. Mezzacappa {\it et~al.}, Physical Review Letters,  submitted  (2000).

\bibitem{tb00}
T.~A. Thompson and A. Burrows, astro-ph/0009449  .

\bibitem{s98}
F.~D. Swesty,  in {\em Stellar Explosions, Stellar Evolution, and Galactic
  Chemical Evolution}, edited by A. Mezzacappa (IoP Publishing, Bristol, 1998),
  pp.\ 539--548.

\bibitem{lw70}
J.~M. LeBlanc and J.~R. Wilson, Astrophysical Journal {\bf 161},  541  (1970).

\bibitem{sy84}
E.~M.~D. Symbalisty, Astrophysical Journal {\bf 285},  729  (1984).

\bibitem{mw99}
A.~I. MacFadyen and S.~E. Woosley, Astrophysical Journal {\bf 524},  262
  (1999).

\bibitem{rj00}
H.-T. Janka and M. Rampp, Astrophysical Journal {\bf 539},  L33  (2000).

\bibitem{ba98}
G. Bazan and W.~D. Arnett, Astrophysical Journal {\bf 496},  316  (1998).

\bibitem{unn99}
H. Umeda, K. Nomoto, and T. Nakamura,  in {\em The First Stars}, edited by A.
  Weiss (Springer, Berlin, 2000).

\bibitem{hlw00}
A. Heger, N. Langer, and S. Woosley, Astrophysical Journal {\bf 528},  368
  (2000).

\bibitem{hlmw00}
A. Heger, K.-H. Langanke, G. Martinez-Pinedo, and S. Woosley, astro-ph/0007412
  .

\bibitem{lmp00}
K.-H. Langanke and G. Martinez-Pinedo, Nuclear Physics {\bf A673},  481
  (2000).

\bibitem{bs98}
A. Burrows and R.~F. Sawyer, Physical Review {\bf C58},  554  (1998).

\bibitem{rplp99}
S. Reddy, M. Prakash, J.~M. Lattimer, and J.~A. Pons, Physical Review {\bf
  C59},  2888  (1999).

\end{thebibliography}
\end{document}